\documentclass{amsproc}
\usepackage{amssymb}

\newtheorem {theorem}{Theorem}

\theoremstyle{definition}
\newtheorem {definition}{Definition}

\theoremstyle{remark}

\begin{document}

\title{The Flux-Across-Surfaces Theorem for a Point Interaction Hamiltonian }

%%%%Author 1

\author{G. Panati}
\address{S.I.S.S.A/I.S.A.S, International School for Advanced Studies,
Via Beirut, 2 - 34014 Trieste, Italy.}
\email{panati@sissa.it}

%%%%Author 2

\author{A. Teta}
\address{Dipartimento di Matematica, Universit\`{a} di Roma ``La
Sapienza'', P.le Aldo Moro, 2 - 00185 Roma, Italy}
\email{teta@mat.uniroma1.it}
\thanks{}


\begin{abstract}
The flux-across-surfaces theorem establishes a fundamental relation in
quantum scattering theory between the asymptotic outgoing state and a
quantity which is directly measured in experiments. We prove it for a
hamiltonian with a point interaction, using the explicit expression for the
propagator. The proof requires only assuptions on the initial state and it
covers also the case of zero-energy resonance. We also outline a different
approach based on generalized eigenfunctions, in view of a possible
extension of the result.
\end{abstract}

\maketitle

\hfill {\em Dedicated to Sergio Albeverio}

\vspace{1cm}

\section{Introduction}

In quantum scattering theory one is concerned with the derivation of \ the
experimentally measurable quantities from the large time asymptotics of the
quantum state. In particular it is 
reasonable to expect that the probability
that a particle crosses the active surface $\Sigma $ of a very far detector
equals the probability that for large times the particle has a momentum in
the cone $C(\Sigma )=\{\lambda \mathbf{x}\in \mathbb{R}^{3}:\mathbf{x}\in
\Sigma ,\lambda \geq 0\}$ generated by the surface $\Sigma $. The precise
mathematical formulation of this flux-across-surfaces (FAS) conjecture is (\cite
{CNS}, see also the discussion in \cite{TD})

\begin{equation}
\lim_{R\rightarrow \infty }\int_{T}^{\infty }\,dt\int_{\Sigma _{R}}\mathbf{j}%
^{\Psi _{t}}\cdot \mathbf{n}\,d\sigma =\int_{C(\Sigma )}|\widehat{\Psi }_{%
\mathrm{out}}(\mathbf{k})|^{2}\,d^{3}k  \label{FAS}
\end{equation}
for any $T\in \mathbb{R}$, where \textbf{\ }$\mathbf{j}^{\Psi
_{t}}:=2\mathrm{Im}%
(\Psi _{t}^{*}\nabla \Psi _{t})$ is the density of probability current
associated to $\Psi _{t}=e^{-iHt}\Psi _{0}$, $\Sigma _{R}=\{\mathbf{x}\in {%
C(\Sigma )}:|\mathbf{x}|=R\}$, $\Psi _{\mathrm{out}}=\Omega _{+}^{-1}\Psi
_{0}$ is the asymptotic outgoing state and $\quad \widehat{\Psi }\quad $
denotes the Fourier transform of $\Psi $. We have chosen units in which $%
\hbar =1$ and $\ m=\frac{1}{2}$.

The proof of (\ref{FAS}) in the free case was given in \cite{DDGZ},
exploiting the explicit form of the free unitary group (see also \cite
{Teufel th} or \cite{TD} for a more direct proof). The interacting case has
been studied in \cite{AmreinZuleta} for\emph{\ }short range and in \cite
{AmreinPearson} for long range potentials respectively. The proof relies on
the basic assumption that the asymptotic outgoing state $\Psi _{\mathrm{out}%
} $ has a Fourier transform with \ compact support not containing the
origin. More recently, a different proof has been given in \cite{TeufelDurr}
for sufficiently smooth potentials, assuming the absence of zero-energy
resonances and requiring $\Psi _{\mathrm{out}}$ in the Schwartz space
$\mathcal{S}(\mathbb{R}^{3})$.

\noindent All these results are obtained under assumptions that avoid the difficulty due to
zero-energy resonances which produce a slower decay of the wave function for
large times and then make problematic the convergence of the left hand side of
(\ref{FAS}).

Here we consider a specific model hamiltonian, i.e. hamiltonian with a point
interaction, and we prove the FAS theorem using the explicit expression for
the propagator, derived in \cite{Sandro}.

\noindent This paper has, in a sense, a pedagogical purpose. It indicates
the possibility to extend the FAS theorem to more general situations. In
particular we show that (\ref{FAS}) holds true in the case of point
interaction even with a zero-energy resonance. The analysis of zero-energy
resonances in the general case of potential scattering will be approached in
a further work (\cite{Panati2}). We also stress that the result is proved assuming only
some regularity on the initial state.

In order to formulate the result we denote by $H_{\alpha ,\mathbf{y}}$ the
Schr\"{o}dinger operator in $L^{2}(\mathbb{R}^{3})$ which corresponds to
one
point interaction placed at $\mathbf{y}\in \mathbb{R}^{3}$ whose strength
is
parametrized by $\alpha \in \mathbb{R}$.

\noindent It is well known that $H_{\alpha ,\mathbf{y}}$ can be constructed
as self-adjoint operator using the standard extension theory (\cite{Albev}). Here we only
recall that the continuous spectrum is purely absolutely continuous and $%
\sigma _{\mathrm{ac}}(H_{\alpha ,\mathbf{y}})=[0,+\infty )$. The point
spectrum is empty if $\alpha \leq 0$ and $\sigma _{\mathrm{p}}(H_{\alpha ,%
\mathbf{y}})=\left\{ -\left( 4\pi \alpha \right) ^{2}\right\} $ if $\alpha
<0 $. For $\alpha =0$ the hamiltonian exhibits a zero-energy resonance.

\noindent The scattering wave functions are 
\begin{equation}
\psi _{\alpha ,\mathbf{y}}(\mathbf{x},\mathbf{k})=e^{i\mathbf{k\cdot x}}+%
\frac{e^{i\mathbf{k\cdot y}}}{(4\pi \alpha -ik)}\frac{e^{ik|\mathbf{x}-%
\mathbf{y}|}}{|\mathbf{x}-\mathbf{y}|}  \label{SWF}
\end{equation}

\noindent Using the generalized eigenfunctions $\Phi _{-}(\mathbf{x},\mathbf{%
k})=\psi _{\alpha ,\mathbf{y}}(\mathbf{x},\mathbf{k})$ and $\Phi _{+}(%
\mathbf{x},\mathbf{k})=$ $\psi _{\alpha ,\mathbf{y}}^{*}(\mathbf{x},-\mathbf{k})
$ one can define two unitary maps $\mathcal{F}_{\pm }:\mathcal{H}%
_{ac}(H_{\alpha ,\mathbf{y}})$ $\rightarrow L^{2}(\mathbb{R}^{3})$ by 

\begin{equation}
(\mathcal{F}_{\pm }f)(\mathbf{k})=\mbox{s-}\!\!\!\!\!\lim_{R\rightarrow +\infty }%
\frac{1}{(2\pi )^{3/2}}\int_{|\mathbf{x}|<R}\Phi _{\pm }^{*}(\mathbf{x},%
\mathbf{k})f(\mathbf{x})\ d^{3}x.  \label{F transforms}
\end{equation}
which spectralise the operator $H_{\alpha ,\mathbf{y}}$ restricted to $%
\mathcal{H}_{ac}$, in the sense that $\mathcal{F}_{\pm }H_{\alpha ,\mathbf{y}}\mathcal{F}_{\pm
}^{-1}$ is a multiplication operator in $L^{2}(\mathbb{R}^{3})$.

The well known relation with the wave operators $\Omega _{\pm }=
\mbox{s-}\!\lim_{t\rightarrow \pm \infty }e^{iH_{\alpha ,\mathbf{y}}t}e^{-iH_{0}t}$
(where $H_{0}=-\Delta $) is expressed by the intertwining properties\emph{\ }

\begin{equation}
\Omega _{\pm }^{-1}=\mathcal{F}^{-1}\mathcal{F}_{\pm }\qquad \text{and\qquad 
}\Omega _{\pm }=\mathcal{F}_{\pm }^{-1}\mathcal{F}.  \label{WOP relations}
\end{equation}

With the above notation, our result is the following.\medskip

\begin{theorem}
\label{Th 1}Let us fix $\Psi _{0}\in \mathcal{S}(\mathbb{R}^{3})\cap
\mathcal{H}%
_{ac}(H_{\alpha ,\mathbf{y}})$. Then $\Psi _{t}:=e^{-iH_{\alpha ,y}t}\Psi
_{0}$ is continuosly differentiable in $\mathbb{R}^{3}\backslash \{0\}$
and
relation (\ref{FAS}) holds true, for every $T\in \mathbb{R}$.
\end{theorem}

\noindent

\noindent In Section 2 we shall give the details of the proof for the more
interesting case $\alpha =0$ and then we outline the procedure for the other
cases. It will also be clear from the proof that the strong assumption $\Psi
_{0}\in \mathcal{S}(\mathbb{R}^{3})$ has been considered only for the sake
of
simplicity; it can be replaced by the assumption that $\Psi _{0}$ is sufficiently
many times differentiable and decays rapidly at infinity.

\noindent In Section 3 we briefly sketch a different proof based on the
generalized eigenfunctions of $H_{\alpha ,\mathbf{y}}$. \ In the case of one
point interaction such method is less satisfactory than the method of the
propagator, since it requires regularity of $\Psi _{\mathrm{out}}$.
Nevertheless it may be  suitable for the extension of the result to
general
potential scattering in presence of zero-energy resonances.

\section{Proof of the theorem}

We shall denote by $w$ the modulus of the vector $\mathbf{w}\in
\mathbb{R}^{3}$
and with $\omega _{\mathbf{w}}:=\frac{\mathbf{w}}{\left| \mathbf{w}\right| }$
the unit vector in the direction defined by $\mathbf{w\neq 0}$.

Without loss of generality we fix $\mathbf{y}=0$ and, since the interaction is not trivial
only in the s-wave (\cite{Albev}), we choose a spherically
symmetric initial state $\Psi _{0}$.

\noindent Using the explicit propagator for $\alpha =0$ (\cite{Sandro}) we
have 
\begin{eqnarray}
\Psi _{t}(\mathbf{x})
&=&\int_{\mathbb{R}^{3}}\frac{e^{i\frac{|\mathbf{x}-%
\mathbf{y}|^{2}}{4t}}}{(4\pi it)^{3/2}}\Psi _{0}(y)\ d^{3}y+\frac{2it}{x}%
\int_{\mathbb{R}^{3}}\ \frac{\Psi _{0}(y)}{y}\frac{e^{i}\frac{\left(
x+y\right)
^{2}}{4t}}{(4\pi it)^{3/2}}\ d^{3}y  \nonumber \\
&=&\frac{\ e^{i\frac{x^{2}}{4t}}}{(2it)^{3/2}}\left[ \frac{1}{(2\pi )^{3/2}}%
\int_{\mathbb{R}^{3}}\Psi _{0}(y)\ \left( e^{-i\frac{\mathbf{x\cdot
y}}{2t}}+%
\frac{2it}{x}\frac{e^{i\frac{x}{2t}y}}{y}\right) \ d^{3}y\right] +R(\mathbf{x%
},t)  \nonumber \\
&=&\frac{\ e^{i\frac{x^{2}}{4t}}}{(2it)^{3/2}}\widehat{\Psi }_{\mathrm{out}%
}\left( \frac{\mathbf{x}}{2t}\right) +R(\mathbf{x},t)\equiv P(\mathbf{x}%
,t)+R(\mathbf{x},t)  \label{Decompo}
\end{eqnarray}
where we used (\ref{WOP relations}) and we have denoted 
\begin{eqnarray}
R(\mathbf{x},t) &=&\frac{e^{i\frac{x^{2}}{4t}}}{(4\pi
it)^{3/2}}\int_{\mathbb{R}%
^{3}}\ e^{-i\frac{\mathbf{x\cdot y}}{2t}}\left( e^{i\frac{y^{2}}{2t}%
}-1\right) \Psi _{0}(y)\ d^{3}y+  \nonumber \\
&+&\frac{\ e^{i\frac{x^{2}}{4t}}}{(2it)^{1/2}x}\frac{1}{(2\pi )^{3/2}}\int_{%
\mathbb{R}^{3}}e^{i\frac{xy}{2t}}\left( e^{i\frac{y^{2}}{2t}}-1\right)
\frac{%
\Psi _{0}(y)}{y}\ d^{3}y  \nonumber \\
&=:&R_{1}(\mathbf{x},t)+R_{2}(\mathbf{x},t)  \label{R}
\end{eqnarray}

\noindent An explicit computation gives

\begin{equation}
\nabla P(\mathbf{x},t)=\frac{i}{2}\frac{\mathbf{x}}{t}P(\mathbf{x},t)+\frac{%
\ e^{i\frac{x^{2}}{4t}}}{(2it)^{3/2}}\nabla _{\mathbf{x}}\widehat{\Psi }_{%
\mathrm{out}}\left(
\frac{\mathbf{x}}{2t}\right)\equiv\frac{i}{2}\frac{\mathbf{x}}%
{t}P(\mathbf{x},t)+\mathbf {Q}(\mathbf{x},t)   \label{grad P}
\end{equation}
Taking (\ref{grad P}) into account one obtains 
\[
\mathbf{j}^{\Psi _{t}}(\mathbf{x},t):=2\mathrm{Im}\left( \Psi _{t}^{*}(\mathbf{%
x},t)\nabla \Psi _{t}(\mathbf{x},t)\right) =\frac{\mathbf{x}}{t}|P(\mathbf{x}%
,t)|^{2}+\mathbf{N}(\mathbf{x},t)
\]
where 
\begin{equation}
\mathbf{N}=2\mathrm{Im}\left( P^{*}\mathbf {Q}+P^{*}\nabla R%
+R^{*}\nabla P+R^{*}\nabla R \right) 
\label{N}
\end{equation}
A change of the integration variable yields 
\begin{eqnarray*}
\lim_{R\rightarrow \infty }\int_{T}^{\infty }\,dt\int_{\Sigma _{R}}|P(%
\mathbf{x},t)|^{2}\frac{\mathbf{x}}{t}\cdot \mathbf{n}\,d\sigma 
&=&\lim_{R\rightarrow \infty }\int_{T}^{\infty }\,dt\int_{\Sigma _{R}}\frac{1%
}{(2t)^{3}}\left| \widehat{\Psi }_{0}\left( \frac{\mathbf{x}}{2t}\right)
\right| ^{2}\frac{\mathbf{x}}{t}\cdot \mathbf{n}\,d\sigma  \\
&=&\int_{C(\Sigma )}|\widehat{\Psi }_{\mathrm{out}}(\mathbf{k})|^{2}\,d^{3}k.
\end{eqnarray*}
We are then reduced to prove that 
\begin{equation}
\lim_{R\rightarrow \infty }\int_{T}^{\infty }\,dt\int_{\Sigma _{R}}\mathbf{N}%
(\mathbf{x},t)\cdot \mathbf{n}\,d\sigma =0  \label{Azzeramento}
\end{equation}
It is convenient to
introduce the following definition.

\begin{definition}
\label{Def}We say that $F:\mathbb{R}^{3}\times {\mathbb{R}\rightarrow
C}^{r}$
 ($r \in {N}$) satisfies the hypothesis $\mathcal{O}(n)$ (for $n\in {N}$), and we write 
$F=\mathcal{O}(n)$, if there exist $T_{0}>0$, $R_{0}>0$ such that 
\[
\sup_{x\geq R_{0},t\geq T_{0}}\left( \frac{x}{t}\right) ^{q}\left| F(\mathbf{%
x},t)\right| \leq C
\]
for each $q\leq n$.  If $F=%
\mathcal{O}(n)$ for every $n\in {N}$ we will write $F=\mathcal{O}(\infty
).$
\end{definition}

\noindent It is easy to check that for every $f\in
\mathcal{S}(\mathbb{R}^{3})$
we have 
\begin{equation}
F(\mathbf{x},t):=\int_{\mathbb{R}^{3}}\ e^{-i\frac{\mathbf{x\cdot
y}}{2t}}f(%
\mathbf{y})\ d^{3}y=\mathcal{O}(\infty ).  \label{Ausil1}
\end{equation}
Likewise,
by iterated integration by parts, one can prove that for every $f\in 
\mathcal{S}(\mathbb{R}^{3})$ 
\begin{equation}
F_{0}(\mathbf{x},t):=\int_{\mathbb{R}^{3}}e^{i\frac{x}{2t}y}\
f(\mathbf{y})\
d^{3}y=\mathcal{O}(2)  \label{Ausil2}
\end{equation}
\begin{equation}
F_{-1}(\mathbf{x},t):=\int_{\mathbb{R}^{3}}e^{i\frac{x}{2t}y}\frac{f(\mathbf{y})%
}{y}\ \ d^{3}y=\mathcal{O}(1)  \label{Ausil3}
\end{equation}
\ 

Using (\ref{Ausil1}), (\ref{Ausil2}) and (\ref{Ausil3}) we can write 
\begin{eqnarray}
P(\mathbf{x},t) &=&\frac{\ e^{i\frac{x^{2}}{4t}}}{(2it)^{3/2}}\widehat{\Psi }%
_{0}\left( \frac{\mathbf{x}}{2t}\right) +\frac{\ e^{i\frac{x^{2}}{4t}}}{%
(2it)^{1/2}x}\frac{1}{(2\pi )^{3/2}}\int_{\mathbb{R}^{3}}\Psi _{0}(y)\
\frac{%
e^{i\frac{x}{2t}y}}{y}\ d^{3}y  \nonumber \\
&=&\frac{\mathcal{O}(\infty )}{t^{3/2}}+\frac{\mathcal{O}(1)}{xt^{1/2}}.
\label{P stima}
\end{eqnarray}
By a direct computation we obtain 
\begin{eqnarray}
\mathbf {Q}(\mathbf{x},t)&=&\frac{\ e^{i\frac{%
x^{2}}{4t}}}{(2it)^{3/2}}\frac{1}{(2\pi )^{3/2}}\left\{ -\frac{i}{2t}\int_{%
\mathbb{R}^{3}}e^{-i\frac{\mathbf{x\cdot y}}{2t}}\mathbf{y}\Psi _{0}(y)\
d^{3}y\right. +  \nonumber \\
&-&\left. i\frac{2t}{x^{2}}\omega
_{\mathbf{x}}\int_{\mathbb{R}^{3}}e^{i\frac{x%
}{2t}y}\ \frac{\Psi _{0}(y)}{y}\ d^{3}y-\frac{1}{x}\omega _{\mathbf{x}}\int_{%
\mathbb{R}^{3}}e^{i\frac{x}{2t}y}\ \Psi _{0}(y)\ d^{3}y\right\}  \nonumber
\\
&=&\frac{1}{t^{5/2}}\mathcal{O}(\infty )+\frac{1}{x^{2}t^{3/2}}\mathcal{O}%
(1)+\frac{1}{xt^{3/2}}\mathcal{O}(2).  \label{grad P stima}
\end{eqnarray}
where we have used (\ref{Ausil1}) and (\ref{Ausil3}) and the inequality 
\begin{equation}
\left| e^{iw}-1\right| \leq \left| w\right| .\qquad  \label{Magic}
\end{equation}

Now we analyse $R_{1}(\mathbf{x},t)$ in (\ref{R}) following the line of \cite{TD}.
For any $q\in {N}$ one has 
\[
\left( \frac{x}{2t}\right) ^{q}e^{-i\frac{\mathbf{x\cdot y}}{2t}%
}=i^{q}\left( \omega _{\mathbf{x}}\cdot \nabla _{y}\right) ^{q}e^{-i\frac{%
\mathbf{x\cdot y}}{2t}} 
\]
and, by iterated integration by parts, we obtain 
\begin{eqnarray*}
\left( \frac{x}{t}\right) ^{q}\left| R_{1}(\mathbf{x},t)\right| &=&\frac{1}{%
(4\pi it)^{3/2}}\left| \int_{\mathbb{R}^{3}}\left[ \left( \omega
_{\mathbf{x}%
}\cdot \nabla _{y}\right) ^{q}e^{-i\frac{\mathbf{x\cdot y}}{2t}}\right] \
\left( e^{i\frac{y^{2}}{2t}}-1\right) \Psi _{0}(y)\ d^{3}y\right| \\
&\leq &\frac{1}{(4\pi it)^{3/2}}\left| \int_{\mathbb{R}^{3}}\left( \omega
_{%
\mathbf{x}}\cdot \nabla _{y}\right) ^{q}\ \left( \left( e^{i\frac{y^{2}}{2t}%
}-1\right) \Psi _{0}(y)\right) \ d^{3}y\right| \leq \frac{C_{q}}{t^{5/2}}
\end{eqnarray*}
where the last inequality follows from an explicit differentiation and from (\ref{Magic}).
 We conclude that 
\begin{equation}
R_{1}(\mathbf{x},t)=\frac{1}{t^{5/2}}\mathcal{O}(\infty )  \label{R1 stima}
\end{equation}
An analogous computation gives 
\begin{eqnarray}
\nabla R_{1}(\mathbf{x},t) &=&\frac{\mathbf{x}}{t}R_{1}(\mathbf{x},t)+\frac{%
e^{i\frac{x^{2}}{4t}}}{(4\pi it)^{3/2}}\frac{-i}{2t}\int_{\mathbb{R}^{3}}\
e^{-i%
\frac{x}{2t}y}\left( e^{i\frac{y^{2}}{2t}}-1\right) \mathbf{y}\Psi _{0}(y)\
d^{3}y  \nonumber \\
&=&\frac{\mathbf{x}}{t^{7/2}}\mathcal{O}(\infty )+\frac{1}{t^{7/2}}\mathcal{O%
}(\infty ).  \label{grad R1 stima}
\end{eqnarray}

We analyse now $R_{2}(\mathbf{x},t)$. Using the inequality (\ref
{Magic}) one easily gets 
\begin{equation}
\left| R_{2}(\mathbf{x},t)\right| \leq \frac{C}{xt^{3/2}}.  \label{In q=0}
\end{equation}
Moreover, using the radial simmetry of $\Psi _{0}$, one has (for $q=1,2$) 
\begin{eqnarray*}
i^{q}\left( \frac{x}{2t}\right) ^{q}R_{2}(\mathbf{x},t) &=&\frac{\ C}{%
xt^{1/2}}\int_{\mathbb{R}^{3}}\left( \frac{d^{q}}{dy^{q}}e^{i\frac{x}{2t}%
y}\right) \left( e^{i\frac{y^{2}}{2t}}-1\right) \frac{\Psi (y)}{y}\ d^{3}y
\\
&=&\frac{\ C}{xt^{1/2}}\int_{0}^{+\infty }\left( \frac{d^{q}}{dy^{q}}e^{i%
\frac{x}{2t}y}\right) \left( e^{i\frac{y^{2}}{2t}}-1\right) \Psi (y)y\ dy.
\end{eqnarray*}
Integrating by parts one obtains 
\begin{equation}
\left| \left( \frac{x}{2t}\right) ^{q}R_{2}(\mathbf{x},t)\right| \leq \frac{%
\ C}{xt^{1/2}}\int_{0}^{+\infty }\left| \frac{d^{q}}{dy^{q}}\left( \left(
e^{i\frac{y^{2}}{2t}}-1\right) \Psi (y)y\right) \right| \ dy\leq \frac{C}{%
xt^{3/2}}  \label{In q}
\end{equation}
By inequalities (\ref{In q=0}) and (\ref{In q}) one has 
\begin{equation}
R_{2}(\mathbf{x},t)=\frac{1}{xt^{3/2}}\mathcal{O}(2)  \label{R2 stima}
\end{equation}
A direct computation yields 
\begin{eqnarray*}
\nabla R_{2}(\mathbf{x},t) &=&\frac{\mathbf{x}}{t}R_{2}(\mathbf{x},t)+\frac{%
e^{i\frac{x^{2}}{4t}}}{(2it)^{1/2}}\frac{1}{(2\pi )^{3/2}}\left\{ -\frac{\omega _{\mathbf{x}}}{%
x^{2}}\int_{\mathbb{R}^{3}}e^{i\frac{x}{2t}y}\left( e^{i%
\frac{y^{2}}{2t}}-1\right) \frac{\Psi (y)}{y}\ d^{3}y\right. \\
&+&\left.
\frac{1}{x}\frac{i}{2t}\int_{\mathbb{R}^{3}}e^{i\frac{x}{2t}y}\left(
e^{i\frac{y^{2}}{2t}}-1\right) \mathbf{y}\Psi _{0}(y)\ d^{3}y\right\} .
\end{eqnarray*}
and, by an argument analogous to the previous one, we obtain 
\begin{equation}
\nabla R_{2}(\mathbf{x},t)=\frac{1}{t^{5/2}}\mathcal{O}(2)+\frac{1}{%
x^{2}t^{3/2}}\mathcal{O}(1)+\frac{1}{xt^{5/2}}\mathcal{O}(2).
\label{grad R2 stima}
\end{equation}

In order to prove (\ref{Azzeramento}) we first observe that 
\begin{eqnarray*}
\lefteqn{P^{*}(\mathbf{x},t)\mathbf {Q}(\mathbf{x},t) =} \\
&=& \mathbf {A}_{0}(\mathbf{x},t)+\frac{%
\mathcal{O}(\infty )}{t^{4}}+\frac{\mathcal{O}(\infty )\mathcal{O}(1)}{%
x^{2}t^{3}}+ \\
&+& \frac{\mathcal{O}(\infty )\mathcal{O}(2)}{xt^{3}}+\frac{\mathcal{O}(1)%
\mathcal{O}(\infty )}{xt^{3}}+\frac{\mathcal{O}(1)\mathcal{O}(2)}{x^{2}t^{2}}%
.  \label{P gradP}
\end{eqnarray*}
where
\begin{eqnarray} 
\mathbf {A}_{0}(\mathbf{x},t) &:=& \frac{1}{(2\pi
)^{3}}\frac{\mathbf{x}}{2tx^{4}}\left| \int_{\mathbb{R}%
^{3}}e^{i\frac{x}{2t}y}\ \frac{\Psi _{0}(y)}{y}\ d^{3}y\right| ^{2}.
\label{A0}
\end{eqnarray}
Since we are interested in the imaginary part of \ (\ref{P gradP}) we can
neglet the term $\mathbf {A}_{0}(\mathbf{x},t)$. For the other terms in (\ref{P gradP}) a direct
application of the dominated convergence theorem yields 
\[
\lim_{R\rightarrow \infty }\int_{T}^{+\infty }\int_{\Sigma _{R}}\left| \mathrm{Im}
\left( P(\mathbf{x},t)\mathbf {Q}(\mathbf{x},t) \right) \right|
R^{2}d\Omega dt=0. 
\]

\noindent Using the estimates (\ref{P stima}), (\ref{grad P stima}), (\ref
{R1 stima}), (\ref{grad R1 stima}), (\ref{R2 stima}) and (\ref{grad R2 stima}%
) we similarly prove that 
\[
\lim_{R\rightarrow \infty }\int_{T}^{+\infty }\int_{\Sigma _{R}}\left| \mathrm{Im}
\left( P^{*}\nabla R+R^{*}\nabla P+R^{*}\nabla R\right) \right|
R^{2}d\Omega dt=0, 
\]
proving the claim for an appropriate $T>0$. Using the invariance by finite
time translations like in \cite{TeufelDurr} we obtain the thesis for every $%
T\in \mathbb{R}$, in the case $\alpha =0$.\bigskip

Now we sketch the proof in the case $\alpha \neq 0$. For $\alpha >0$, using
the explicit form of the propagator, one obtains 
\begin{eqnarray*}
\Psi _{t}(\mathbf{x}) &=&\frac{\ e^{i\frac{x^{2}}{4t}}}{(4\pi it)^{3/2}}
\int_{\mathbb{R}^{3}}\Psi _{0}(y)\left( e^{-i\frac{%
\mathbf{x\cdot y}}{2t}}+\frac{1}{4\pi \alpha -i\frac{x}{2t}}\frac{e^{i\frac{x%
}{2t}y}}{y}\right) d^{3}y +\sum_{j=1}^{3}R_{j}(\mathbf{x},t) \\
&=&\frac{\ e^{i\frac{x^{2}}{4t}}}{(2it)^{3/2}}\widehat{\Psi }_{\mathrm{out}%
}\left( \frac{\mathbf{x}}{2t}\right) +\sum_{j=1}^{3}R_{j}(\mathbf{x},t)
\end{eqnarray*}
where $R_{1}$ and $R_{2}$ are given by (\ref{R}) and 
\[
R_{3}(\mathbf{x},t)\!:=\!\frac{-2\alpha e^{i\frac{x^{2}}{4t}}}{(4\pi it)^{1/2}x}%
\int_{\mathbb{R}^{3}}\!d^{3}y\ e^{i\frac{x}{2t}y}\frac{\Psi _{0}(y)}{y}%
\int_{0}^{+\infty }\!\!\!\!du\ e^{-4\pi \alpha u}\left( e^{\frac{i}{4t}%
(u^{2}+y^{2}+2uy)}-1\right) e^{\frac{i}{2t}ux}.
\]

\noindent

\noindent All the estimates proved for $\alpha=0$ hold true in the present case.
 Moreover, concerning the leading term it is easy to see that 
\begin{equation}
\left| \frac{\ e^{i\frac{x^{2}}{4t}}}{(2it)^{3/2}}\widehat{\Psi }_{\mathrm{%
out}}\left( \frac{\mathbf{x}}{2t}\right) \right| \leq \frac{C}{t^{3/2}}.
\label{P stima NR}
\end{equation}
We stress that the estimate (\ref{P stima NR}) holds true only for 
$\alpha \neq 0$.
Using the radial symmetry of $\Psi _{0\text{ }}$ and applying Fubini's
theorem we can write 
\begin{eqnarray*}
\widetilde{R}_{3}(\mathbf{x},t) &:=&\int_{\mathbb{R}^{3}}d^{3}y\
e^{i\frac{%
x}{2t}y}\frac{\Psi _{0}(y)}{y}\int_{0}^{+\infty }du\ e^{-4\pi \alpha
u}\left( e^{\frac{i}{4t}(u^{2}+y^{2}+2uy)}-1\right) e^{\frac{i}{4t}2ux} \\
&=& 4\pi \int_{0}^{+\infty }du\ \int_{0}^{+\infty }dy\ e^{i\frac{x}{2t}\left(
y+u\right) }y\Psi _{0}(y)e^{-4\pi \alpha u}\left( e^{\frac{i}{4t}%
(u+y)^{2}}-1\right) \\
&=& 4 \pi \int_{0}^{+\infty }du\ \int_{u}^{+\infty }dw\ e^{i\frac{x}{2t}%
w}(w-u)\Psi _{0}(w-u)e^{-4\pi \alpha u}\left( e^{\frac{i}{4t}w^{2}}-1\right)
\end{eqnarray*}
where we used the change of integration variable $w=y+u$. The integration
domain is 
\[
D=\left\{ (u,w)\in \mathbb{R}_{+}^{2}:u\in [0,+\infty ),w>u\right\}
=\left\{
(u,w)\in \mathbb{R}_{+}^{2}:w\in [0,+\infty ),u<w\right\} 
\]
so 
\begin{eqnarray*}
\widetilde{R}_{3}(x,t) &=& 4 \pi \int_{0}^{+\infty }dw\ \int_{0}^{w}du\ e^{i\frac{x%
}{2t}w}\left( e^{\frac{i}{4t}w^{2}}-1\right) (w-u)\Psi _{0}(w-u)e^{-4\pi
\alpha u} \\
&=& 4 \pi \int_{0}^{+\infty }dw\ \ e^{i\frac{x}{2t}w}\left( e^{\frac{i}{4t}%
w^{2}}-1\right) e^{-4\pi \alpha w}\int_{0}^{w}ds\ s\Psi _{0}(s)e^{4\pi
\alpha s}.
\end{eqnarray*}
Using the fact that the function 
\[
\varphi (w):=e^{-4\pi \alpha w}\int_{0}^{w}ds\ s\Psi _{0}(s)e^{4\pi \alpha
s} 
\]
satisfies 
\begin{equation}
\lim_{w\rightarrow +\infty }w^{n}\varphi (w)=0\qquad (n\in {N})
\label{Phi asympto}
\end{equation}
we can integrate by parts showing that $\widetilde{R}_{3}(x,t)=\frac{%
1}{t}\mathcal{O}(2)$. Then we conclude that 
\begin{equation}
R_{3}(\mathbf{x},t)=\frac{\mathcal{O}(2)}{xt^{3/2}}.  \label{R3 stima}
\end{equation}
Following the same line we obtain the corrisponding estimate for $\nabla R_{3}(\mathbf{x},t)$
\begin{equation}
\nabla R_{3}(\mathbf{x},t)=\frac{\mathcal{O}(2)}{t^{5/2}} + 
\frac{\mathcal{O}(2)}{x^{2}t^{3/2}} + \frac{\mathcal{O}(2)}{x t^{5/2}}.
\label{grad R3 stima}
\end{equation}
The previous estimates (\ref{P stima NR}), (\ref{R3 stima}) and (\ref{grad R3 stima}) allow us to
use dominated convergence and then to obtain the result in the case $%
\alpha >0$.\bigskip

Finally, in the case $\alpha <0$ the propagator can be written in the form 
\begin{eqnarray*}
\Psi _{t}(\mathbf{x})&=&\left( e^{-iH_{0}t}\Psi _{0}\right) (\mathbf{x}%
)-e^{i(4\pi \alpha )^{2}t}\Psi _{\alpha
}(\mathbf{x})\int_{\mathbb{R}^{3}}\Psi
_{\alpha }^{*}(y)\Psi _{0}(y)\ d^{3}y+ \\
&+&\!\int_{\mathbb{R}^{3}}\!\!\frac{\Psi
_{0}(y)}{y}\frac{e^{i}\frac{\left(
x+y\right) ^{2}}{4t}}{(2it)^{1/2}x}d^{3}y-4\pi \alpha \frac{2it}{x}\int_{%
\mathbb{R}^{3}}d^{3}y\ \frac{\Psi _{0}(y)}{y}\int_{0}^{+\infty }\!\!\!du\
e^{4\pi
\alpha u}\frac{e^{i}\frac{\left( u+x+y\right) ^{2}}{4t}}{(4\pi it)^{3/2}}
\end{eqnarray*}
where $\Psi _{\alpha }$ is the eigenfunction relative to the eigenvalue $%
\lambda _{\alpha }=-\left( 4\pi \alpha \right) ^{2}$. The second term is
identically zero due to the assumption $\Psi _{0}\in \mathcal{H}_{\mathrm{ac}%
}(H_{\alpha ,\mathbf{y}})=\mathcal{H}_{\mathrm{p}}(H_{\alpha ,\mathbf{y}%
})^{\perp }$ and all the remaining terms can be treated exactly as in the
case $\alpha >0$, so we omit the details. \bigskip

\noindent \textbf{Remark 1. } In the proof of Theorem 1 for $\alpha=0$ we have
 estimated the {\em absolute value} 
of the terms in parenthesis in (\ref{N}), except for $\mathbf {A}_{0}$ 
(see (\ref{A0})), which is real and then it doesn't contribute to $\mathbf{N}$.

\noindent This is a crucial point since, in general, one has 
\begin{equation}
\lim_{R\rightarrow \infty }\int_{T}^{\infty }\,dt\int_{\Sigma _{R}}
\left| \mathbf {A}_{0}\cdot \mathbf{n} \right| \,d\sigma =+\infty 
\label{divergence}
\end{equation}

\noindent (choose e.g. $\Psi_{0}(y)=e^{-y}$).
 
\noindent The divergent limit (\ref{divergence}) is a consequence of 
the slower decay of the wave function in presence of a zero-energy 
resonance. This means that, in such case, one cannot hope to prove 
the theorem by simply estimating the absolute value of $\mathbf{N}$,
unless one assumes the pseudo-orthogonality condition
 $\Psi_{0} \in \mathcal{W}$ (see Remark 2). 
\noindent On the other hand this difficulty doesn't arise for 
$\alpha \neq 0$. In fact the term $\mathbf {A}_{0}$ is now replaced by

\begin{equation}
\mathbf {A}_{\alpha}(\mathbf{x},t) = \frac{1}{t^{3}}\left|
\int_{\mathbb{R}%
^{3}}e^{i\frac{x}{t}y}\ \frac{\Psi _{0}(y)}{y}\ d^{3}y\right| ^{2}
\frac{1}{\alpha+i\frac{x}{t}}
\left( \frac{d}{dx}\frac{1}{\alpha-i\frac{x}{t}} \right)
\omega_{\mathbf{x}}
\label{Aalpha}
\end{equation}

\noindent which is easily estimated taking the absolute value.

\section{The method of generalized eigenfunctions}

In this section we outline a proof of the FAS\ conjecture based on the
generalized eigenfunctions of $H_{\alpha ,\mathbf{y}}$ and we compare it
with the result obtained in the previous section. This tecnique was
previosly used in \cite{TeufelDurr}. Here we generalize the method in
order to allow for the presence of zero-energy resonances.

\begin{theorem}
\label{Th2}Let us fix $\Psi _{\mathrm{out}}\in
\mathcal{S}(\mathbb{R}^{3})$.
Then $%
\Psi _{t}:=e^{-iH_{\alpha ,\mathbf{y}}t}\Omega _{+}\Psi _{\mathrm{out}}$ is
continuosly differentiable in $\mathbb{R}^{3}\backslash \{0\}$ and (\ref
{FAS}) holds true for every $T\in \mathbb{R}$.
\end{theorem}

\noindent\textbf{Proof.} Let be $\Psi _{0}=\Omega _{+}\Psi _{\mathrm{out}}$. Using
the properties of $\mathcal{F}_{+}$ and (\ref{WOP relations}) we obtain 
\begin{eqnarray*}
\Psi _{t}(\mathbf{x})&=&\frac{1}{(2\pi
)^{\frac{3}{2}}}\int_{\mathbb{R}^{3}}e^{-ik^{2}t}\widehat{%
\Psi }_{\mathrm{out}}(\mathbf{k})\Phi _{+}(\mathbf{x},\mathbf{k})\,d^{3}k \\
&=&\frac{1}{(2\pi
)^{\frac{3}{2}}}\int_{\mathbb{R}^{3}}e^{-ik^{2}t}\widehat{%
\Psi }_{\mathrm{out}}(\mathbf{k})e^{i\mathbf{k\cdot x}}\,d^{3}k+\frac{1}{%
(2\pi )^{\frac{3}{2}}}\int_{\mathbb{R}^{3}}e^{-ik^{2}t}\frac{\widehat{\Psi
}_{\mathrm{%
out}}(\mathbf{k})}{4\pi \alpha +ik}\frac{e^{-ikx}}{x}d^{3}k \\
&=: &a(\mathbf{x},t)+b(\mathbf{x},t)\,.
\end{eqnarray*}
\noindent The density current is 
\begin{equation}
\mathbf{j}^{\Psi _{t}}=\mathrm{Im}(a^{*}\nabla a+a^{*}\nabla b+b^{*}\nabla
a+b^{*}\nabla a).\,  \label{jt}
\end{equation}
\noindent The first term $\mathbf{j}_{0}=\mathrm{Im}(a^{*}\nabla a)$
corresponds to the free evolution of $\Psi _{\mathrm{out}}$, so using the
free flux-across-surfaces theorem \cite{DDGZ} one has 
\[
\lim_{R\rightarrow \infty }\int_{T}^{+\infty }dt\int_{\Sigma _{R}}\mathbf{j}%
_{0}(\mathbf{x},t)\mathbf{\cdot n}\,d\sigma =\int_{C(\Sigma )}|\widehat{\Psi 
}_{\mathrm{out}}(\mathbf{k})|^{2}\,d^{3}k\,.
\]
\noindent It remains to show that 
\begin{equation}
\lim_{R\rightarrow \infty }\int_{T}^{+\infty }dt\int_{\Sigma _{R}}|\mathbf{j}%
_{1}(\mathbf{x},t)\cdot \mathbf{n}|\,d\sigma =0\,  \label{j1}
\end{equation}
where $\mathbf{j}_{1}:=\mathrm{Im}(a^{*}\nabla b+b^{*}\nabla
a+b^{*}\nabla b)$.

\noindent In order to prove (\ref{j1}) we need estimates on $a,b$ and their
gradients. In the notation of the previous section (see Definition \ref{Def}%
), one has

\noindent

\begin{equation}
a(\mathbf{x},t)=\frac{\mathcal{O}(\infty )}{t^{3/2}}\qquad \text{and\qquad }%
\nabla a(\mathbf{x},t)=\frac{\mathcal{O}(\infty )}{t^{3/2}}.  \label{a}
\end{equation}

\noindent Concerning $b$,$\nabla b$ we distinguish between the cases $\alpha
\neq 0$ and $\alpha =0$.

\noindent \textbf{Case I. }In the case $\alpha \neq 0$ \ we use a stationary
phase tecnique, following \cite{TeufelDurr}. Posing 
\begin{equation}
\chi (k)\equiv \frac{k^{2}t+kx}{t+x}\qquad \text{and\qquad }\omega \equiv t+x
\label{omega}
\end{equation}
and denoting with $^{\prime }$ the derivation respect to $k$, one has

\begin{eqnarray}
\left| b(\mathbf{x},t)\right|  &=&\frac{1}{(2\pi )^{\frac{3}{2}}}\left|
\int_{\mathbb{R}^{3}}e^{i\omega \chi (k)}\widehat{\Psi
}_{\mathrm{out}}(\mathbf{%
k})\frac{1}{4\pi \alpha +ik}\frac{1}{x}d^{3}k\right|   \nonumber
\label{partint} \\
&=&\frac{1}{(2\pi )^{\frac{3}{2}}}\frac{1}{x}\left| \int \frac{1}{\omega
\chi ^{\prime }}\left[ \frac{d}{dk}e^{-i\omega \chi }\right] \widehat{\Psi }%
_{\mathrm{out}}(\mathbf{k})\frac{1}{4\pi \alpha +ik}k^{2}\,dk\,d\Omega _{%
\mathbf{k}}\right|   \nonumber
\end{eqnarray}
Integrating by parts and observing that the boundary term vanishes for $k=0$
since $\chi ^{\prime }(k)\geq \min (1,2k)$, we obtain 
\begin{eqnarray}
\sup_{\mathbf{x}\in S_{R}}\left| b(\mathbf{x},t)\right|  &\leq &\frac{1}{%
R(R+t)}\int \left| \frac{d}{dk}\left[ \frac{1}{\chi ^{\prime }}\widehat{\Psi 
}_{\mathrm{out}}(\mathbf{k})\frac{1}{4\pi \alpha +ik}k^{2}\right] \right|
\,dk\,d\Omega _{\mathbf{k}}  \nonumber \\
&\leq &\frac{C}{R(R+t)}  \label{integral}
\end{eqnarray}

\noindent The function $b(x,t)$ is continuosly differentiable in
$\mathbb{R}%
^{3}\backslash \{0\}$ and for every $\mathbf{x}\neq 0$ one has

\begin{equation}
(\nabla b)(\mathbf{x},t)=-\frac{1}{(2\pi )^{\frac{3}{2}}}\frac{\omega _{%
\mathbf{x}}}{x}\int e^{-i(k^{2}t+kx)}\frac{\widehat{\Psi }_{\mathrm{out}}(%
\mathbf{k})}{4\pi \alpha +ik} \left(\frac{1}{x}+ik \right)\,d^{3}k.
\label{grad b expression}
\end{equation}
The first term is given by $-\frac{\omega _{\mathbf{x}}}{x}b(\mathbf{x},t)$
and the second term is similar to $b(\mathbf{x},t)$ when we replace $%
\widehat{\Psi }_{\mathrm{out}}(\mathbf{k})$ by $k\widehat{\Psi }_{\mathrm{out%
}}(\mathbf{k})$. We obtain

\begin{equation}
\sup_{\mathbf{x}\in S_{R}}\left| \nabla b(\mathbf{x},t)\right| \leq \frac{c}{%
R(R+t)}\,.  \label{grad b}
\end{equation}
Using (\ref{a}), (\ref{integral}) and (\ref{grad b}) it is now easy to
verify (\ref{j1}) and then to prove the thesis.

\noindent \textbf{Case II. }In the case $\alpha =0$ we proceed in a slightly
different way. Denoting $\Psi (k):=\int\nolimits_{S^{2}}\widehat{\Psi }_{%
\mathrm{out}}(\mathbf{k})d\Omega _{\mathbf{k}}$ we write

\[
b(x,t)=\frac{c}{x}\int_{0}^{+\infty }\ e^{-ik^{2}t-ixk}\ k\Psi (k)\ dk.
\]
In order to isolate the contribution arising from the value of $\widehat{%
\Psi }_{\mathrm{out}}$ in zero, we write $b(x,t)=b_{1}(x,t)+b_{2}(x,t)$ 
where
\begin{eqnarray*}
b_{1}(x,t) &:=&\frac{c}{x}\int_{0}^{+\infty }\ e^{-ik^{2}t-ixk}\ k\left( \Psi
(k)-\Psi (0)e^{-k^{2}}\right) \ dk \\
b_{2}(x,t) &:=&\frac{c}{x}\Psi (0)\int_{0}^{+\infty }\ e^{-ik^{2}t-ixk}\ k\
e^{-k^{2}}\ dk
\end{eqnarray*}

The term $b_{1}(x,t)$ can be dealt with by the same technique used in the 
case $\alpha \neq 0$, so we analyse the term $b_{2}(x,t).$ Using the formula 
(valid for $\mathrm{Re}\xi >0,\eta \in {C}$)
\[
\int_{0}^{+\infty }\ w\exp \left( -\xi w^{2}-\eta w\right) dw=-\frac{1}{2}%
\sqrt{\frac{\pi }{\xi }}\frac{\partial }{\partial \eta }\left[ \exp \left( 
\frac{\eta ^{2}}{4\xi }\right) \mathrm{erfc}\left( \frac{\eta }{2\sqrt{\xi }}%
\right) \right]
\]
one obtains that (posing $\xi :=1-it$ and $R=|x|$)

\[
b_{2}(x,t)=-\frac{1}{4}\frac{c}{R\ \xi ^{3/2}}\left( iR\sqrt{\pi }e^{-\frac{%
R^{2}}{4\xi }}-2\sqrt{\xi }-iR\sqrt{\pi }e^{-\frac{R^{2}}{4\xi }}\mathrm{erf}%
\left( \frac{iR}{2\sqrt{\xi }}\right) \right) .
\]
Then we conclude that 
\begin{equation}
\left| b(x,t)\right| \leq \frac{C_{1}}{R(R+t)}+\frac{C_{2}}{R\ t}.
\label{betastima2}
\end{equation}
Concerning the gradient we have 
\begin{eqnarray}
\nabla b_{2}(x,t) &=&-\frac{c}{x^{2}}\ \omega _{x}\int_{0}^{+\infty }\
e^{-ik^{2}t-ixk}\ ke^{-k^{2}}\ dk+  \nonumber \\
&-&i\frac{c}{x}\ \omega _{x}\int_{0}^{+\infty }\ e^{-ik^{2}t-ixk}\
k^{2}e^{-k^{2}}\ dk  \label{grad b2 expr}
\end{eqnarray}

\noindent From (\ref{grad b2 expr}) one easily deduces the estimate 
\begin{equation}
\left| \nabla b(x,t)\right| \leq \frac{C_{1}}{R(R+t)}+\frac{C_{2}}{R^{2}t}.
\label{gradbetastima2}
\end{equation}
Estimates (\ref{betastima2}) and (\ref{gradbetastima2}) allow us to apply
the dominated convergence theorem, and then to prove (\ref{j1}).\textsc{%
\bigskip }

\noindent \textbf{Remark 2. }Using (\ref{WOP relations}), (\ref{F transforms}%
)\ and (\ref{SWF}) one obtains that 
\begin{eqnarray*}
\widehat{\Psi }_{\mathrm{out}}(\mathbf{k}) &=&\left( \mathcal{F}_{+}\Psi
_{0}\right) (\mathbf{k})=\int_{\mathbb{R}^{3}}\Phi
_{+}(\mathbf{x},\mathbf{k}%
)^{*}\Psi _{0}(\mathbf{x})\ d^{3}x \\
&=&\int_{\mathbb{R}^{3}}\left( e^{-i\mathbf{k\cdot x}}+\frac{1}{(4\pi
\alpha
-ik)}\frac{e^{ikx}}{x}\right) \Psi _{0}(\mathbf{x})\ d^{3}x \\
&=&\widehat{\Psi }_{0}(\mathbf{k})+\frac{1}{4\pi \alpha
-ik}\int_{\mathbb{R}%
^{3}}\frac{e^{ikx}}{x}\Psi _{0}(\mathbf{x})\ d^{3}x.
\end{eqnarray*}
\noindent The above  expression shows that, in presence of zero-energy
resonances
(i.e. for $\alpha =0$), and if $\widehat{\Psi }_{0}$ is regular,
the asymptotic outgoing state has a singularity in the origin of
the momentum space, unless
the initial state $\Psi _{0}$ belongs to the linear subspace 
\[
\mathcal{W}=\left\{ \Psi \in
L^{2}(\mathbb{R}^{3}):\int_{\mathbb{R}^{3}}\frac{1}{x}%
\Psi (\mathbf{x})\ d^{3}x=0\right\} 
\]

\noindent We underline that the set $\mathcal{W}$ is not closed in
$L^{2}(\mathbb{R}^{3})$.
Note that, if $\Psi_{0} \in \mathcal{W}$, then the problematic term $\mathbf {A}_{0}$
can be estimated taking the absolute value.
The condition $\Psi _{0}\in \mathcal{W}$ can be read as a
condition of pseudo-orthogonality between $\Psi _{0}$ and the resonance
function $\Psi _{\mathrm{res}}(\mathbf{x})=\frac{1}{|\mathbf{x}|}\in L_{%
\mathrm{loc}}^{2}(\mathbb{R}^{3}).$ 
\noindent It is then clear that an assumption on the
smoothness of $\widehat{\Psi }_{\mathrm{out}}$ is  related to this 
condition of pseudo-orthogonality on the initial state. 
\noindent Since in Theorem \ref{Th 1} we don't need such restrictive assumption on $\Psi _{0}$
we conclude that the method of the propagator allows a better result.
Nevertheless the method cannot be extended to more general situations while,
on the other hand, \ this seems to be the case for the method of generalized
eigenfunctions.\bigskip

\noindent \textbf{Acknowledgements.} We are indebted to prof. G.
Dell'Antonio for many helpful discussions and comments. We also thank prof.
D. D\"{u}rr and dr. S. Teufel for some stimulating discussions during the
preparation of this paper.

\end{document}